\documentclass[aps,prd,superscriptaddress,amsmath,amssymb,twocolumn,fleqn,nofootinbib]{revtex4-1} 

\usepackage{xcolor}
\usepackage{graphicx}
\usepackage{colordvi}
\usepackage{mathtools}
\usepackage{empheq}
\usepackage{bm}%

\usepackage{floatrow}
\usepackage{braket}
\usepackage{soul}

\usepackage{amssymb}
\usepackage{amsmath}

\usepackage{hyperref}
\def\be{\begin{equation}}
\def\ee{\end{equation}}
\def\ba{\begin{eqnarray}}
\def\ea{\end{eqnarray}}

\begin{document}

\title{Unified framework for Early Dark Energy from $\alpha$-attractors}

\author{Matteo Braglia}\email{matteo.braglia2@unibo.it}
\affiliation{Dipartimento di Fisica e Astronomia, Alma Mater Studiorum
Universit\`a di Bologna, \\
Via Gobetti, 93/2, I-40129 Bologna, Italy}
\affiliation{INAF/OAS Bologna, via Gobetti 101, I-40129 Bologna, Italy}
\affiliation{INFN, Sezione di Bologna, Via Berti Pichat 6/2, I-40127 Bologna, Italy}

\author{William T. Emond}\email{william.emond@nottingham.ac.uk}
\affiliation{School of Physics and Astronomy, University of Nottingham, University Park, Nottinham NG7 2 RD, United Kingdom}

\author{Fabio Finelli}\email{fabio.finelli@inaf.it}
\affiliation{INAF/OAS Bologna, via Gobetti 101, I-40129 Bologna, Italy}
\affiliation{INFN, Sezione di Bologna, Via Berti Pichat 6/2, I-40127 Bologna, Italy}

\author{A. Emir G\"{u}mr\"{u}k\c{c}\"{u}o\u{g}lu}\email{emir.gumrukcuoglu@port.ac.uk}
\affiliation{Institute of Cosmology and Gravitation, University of Portsmouth, Dennis Sciama Building, Portsmouth PO1 3FX, United Kingdom}

\author{Kazuya Koyama}\email{kazuya.koyama@port.ac.uk}
\affiliation{Institute of Cosmology and Gravitation, University of Portsmouth, Dennis Sciama Building, Portsmouth PO1 3FX, United Kingdom}

\date{\today}
\begin{abstract}
One of the most appealing approaches to
ease the Hubble tension 
is the inclusion of an early dark energy (EDE) component that adds energy to the Universe in a narrow redshift window around the time of recombination and dilutes faster than radiation afterwards. In this paper, we analyze EDE in the framework of $\alpha$-attractor models. As well known, the success in alleviating the Hubble tension crucially depends on the shape of the energy injection. We show how different types of energy injections can be obtained, thanks to the freedom in choosing the functional form of the potential inspired by  $\alpha$-attractor models. To confirm our intuition we perform an MCMC analysis for three representative cases and find indeed that $H_0$ is significantly larger than in $\Lambda$CDM like in other EDE models.
Unlike axion-driven EDE models with super Planckian decay constant, the curvature of the potential in the EDE  models required by the data is natural in the context of recent theoretical developments in $\alpha$-attractors.
\end{abstract}

\pacs{Valid PACS appear here}
\maketitle
\section{Introduction}
Recent low-redshifts distance-ladder measurements suggest a larger Hubble constant $H_0$ than the one determined from cosmic microwave background (CMB) data \cite{Verde:2019ivm}. The value of $H_0$ inferred from the latest \emph{Planck} 2018 data, $H_0 = (67.36 \pm 0.54)$ km s$^{-1}$Mpc$^{-1}$ \cite{Aghanim:2018eyx},  appears to be in a 4.4$\sigma$ tension with the most recent distance-ladder measurement from the SH0ES team \cite{Riess:2019cxk}, $H_0 = (74.03 \pm 1.42)$ km s$^{-1}$Mpc$^{-1}$ which is determined by using type Ia supernovae (SNe Ia) as standard candles \cite{Riess:2018byc}.  Other low-redshift methods to determine $H_0$, such as from strong-lensing time delay \cite{Wong:2019kwg} or from calibrating SNe Ia by the tip of the red giant branch \cite{Freedman:2019jwv,Yuan:2019npk,Freedman:2020dne}, also point  to a higher $H_0$ derived from the CMB. 
In absence of unknown systematics, new physics seems necessary to solve this $H_0$ tension \cite{Knox:2019rjx}.

A common approach to model building consists of increasing the expansion rate at redshifts around matter-radiation equality in order to shrink the comoving sound horizon at baryon drag $r_s$, which results in a higher $H_0$ inferred from CMB\footnote{For attempts to late time solutions, see e.g. Refs.~\cite{DiValentino:2016hlg,Keeley:2019esp,Benevento:2020fev,Alestas:2020mvb}  } \cite{Bernal:2016gxb,Aylor:2018drw}. 
The addition of light relics  is a typical example of physics that helps ease the tension by changing  the early-time dynamics of the Universe \cite{Lancaster:2017ksf,Buen-Abad:2017gxg,DEramo:2018vss,Kreisch:2019yzn,Blinov:2019gcj}. Modified gravity models also lead  to interesting solutions to the $H_0$ tension \cite{Umilta:2015cta,Ballardini:2016cvy,Lin:2018nxe,Rossi:2019lgt,Sola:2019jek,Zumalacarregui:2020cjh,Ballesteros:2020sik,Braglia:2020iik,Ballardini:2020iws}.

However, given  the success of the $\Lambda$ Cold Dark Matter ($\Lambda$CDM) concordance model in fitting CMB anisotropies, the early time deviation from it must be minimal. To this end, the Early Dark Energy (EDE) scenario is perhaps the most minimal modification to the $\Lambda$CDM background dynamics that substantially alleviates the $H_0$ tension. In this model, first proposed in \cite{Poulin:2018cxd}, a very light scalar field $\phi$ is frozen by Hubble friction during the radiation era, acting as DE with an equation of state $w_{\rm EDE}\equiv P_{\rm EDE}/\rho_{\rm EDE}=-1$ and contributing negligibly to the energy budget of the Universe. Eventually, when the Hubble rate becomes smaller than its effective mass $\partial^2 V(\phi)/\partial\phi^2$, the scalar field quickly rolls down its potential and oscillates around its minimum, its energy diluting faster than radiation. This results in a very sharp energy injection into the cosmic fluid, that for a suitable value of the mass of $\phi$ occurs around the epoch of matter-radiation equality equivalence, successfully lowering $r_s$. Since the seminal work  Ref.~\cite{Poulin:2018cxd}, a substantial effort has been 
made in building new models of EDE \cite{Agrawal:2019lmo,Alexander:2019rsc,Lin:2019qug,Smith:2019ihp,Niedermann:2019olb,Berghaus:2019cls,Sakstein:2019fmf,Kaloper:2019lpl,Gonzalez:2020fdy,Niedermann:2020dwg} and testing their predictions against larger datasets 
\cite{Hill:2020osr,Chudaykin:2020acu}.

In this work, we consider EDE in the framework of $\alpha$-attractors \cite{Kallosh:2013hoa,Kallosh:2013yoa,Galante:2014ifa}, in which the potential for the EDE scalar field is given by 
\begin{equation}
\label{eq:potential1}
    V(\phi)= f^2\left[\tanh(\phi/\sqrt{6 \alpha} M_\textup{pl})\right].
\end{equation} 
 This potential arises naturally by turning a non-canonical kinetic pole-like term of the form $[\alpha /(1 - \varphi^2/6 M_\textup{pl}^2)^2] (\partial \varphi)^2/2$ into a canonical one, that is of the form $(\partial \varphi)^2/2$. Through the field redefinition $\phi = \sqrt{6 \alpha} M_\textup{pl} \tanh^{-1}(\varphi/\sqrt{6}M_\textup{pl})$, the kinetic term becomes canonical and the potential acquires the $\tanh(\phi/\sqrt{\alpha}M_\textup{pl})$ dependence. Due to this field redefinition, the potential flattens to a plateau at large values of $\phi$.
$\alpha$-attractor models were first introduced in the context of inflation, 
with predictions for the spectral index 
$n_s$ and tensor-to-scalar ratio $r$, 
largely independent  of the specific functional form of $V(\phi)$, hence the name ``attractors". 
In the context of dark energy, $\alpha$-attractor models with an energy scale far below the one used in inflation were considered in \cite{Linder:2015qxa,Garcia-Garcia:2018hlc,Cedeno:2019cgr}. An interesting connection between dark energy and inflation for $\alpha$-attractor models has also been investigated in  \cite{Dimopoulos:2017zvq,Akrami:2017cir}.

In our EDE proposal, however, the shape of the potential away from the plateau and around its minimum is crucial, as it regulates the shape of the energy injection. One of the \emph{attractive} features of $\alpha$-attractor models with the potential in Eq.~\eqref{eq:potential1}, is that they can easily accommodate various types of energy injection. Indeed, we will show that, depending on the functional form of $V(\phi)$, a smooth or oscillating energy injection can be produced,  reproducing results of  representative earlier works in the field in a single framework \cite{Poulin:2018cxd,Agrawal:2019lmo,Lin:2019qug}. 

This paper is organized as follows. In Sec.~\ref{sec:background}, we describe the background evolution of the model and compare it to existing EDE models, focusing on the shape of the energy injection. We confirm the capability of our model to alleviate the $H_0$ tension by performing an MCMC analysis in Sec.~\ref{sec:MCMC} and comment on our results in Sec.~\ref{sec:results}. We end in the conclusions Sec.~\ref{sec:conclusions}.

\section{Background evolution and energy injection}~\label{sec:background}
Our model is described by the following Lagrangian
\begin{equation}
\label{eq:Lagrangian}
\mathcal{L}=\sqrt{-g}
\left[\frac{M_\textup{pl}^2}{2}R-\frac{\left(\partial\phi\right)^2}{2}-V(\phi)\right] + \mathcal{L}_m,
\end{equation}
where $\mathcal{L}_m$ is the Lagrangian for matter (including baryons, CDM, photons and neutrinos)
and the potential is:
\begin{equation}
\label{eq:potential2}
V(\phi)=\Lambda+V_0
\frac{(1+\beta)^{2n} \tanh\left(\phi/\sqrt{6\alpha}M_\textup{pl}\right)^{2 p}}
{\left[1+ \beta \tanh\left(\phi/\sqrt{6\alpha}M_\textup{pl}\right)\right]^{2n}} \,,
\end{equation}
where  $V_0,\,p,\,n$, $\alpha$ and $\beta$ are constants.         
The potential corresponds to the simple form $V(x) = \Lambda + 
\tilde{V}_0 x^{2 p}/(1+\tilde{\beta}x)^{2 n}$ for the field $x=\phi/\sqrt{6\alpha}M_\textup{pl}$ with
a pole-like kinetic term and has an offset with respect to Refs.~\cite{Linder:2015qxa,Garcia-Garcia:2018hlc}\footnote{Note that in 
our setting the field rolls towards the minimum of the potential in $\phi=0$ and not towards infinity 
as in \cite{Linder:2015qxa,Garcia-Garcia:2018hlc}.}, 
which is admitted in the dark-energy 
context with $\alpha$-attractors \cite{Akrami:2017cir}.  
We have inserted the normalization factor of $(1+\beta)^{2n}$ 
to ensure the same normalization of the
plateau at large $\phi>0$  for every choice of $(p,\,n)$. 
For  definiteness we will consider $\beta=1$ in the following and we will use a rescaled
scalar field $\Theta\equiv \phi/(\sqrt{6 \alpha} M_\textup{pl})$ when useful. Note that we have added an offset in the potential \eqref{eq:potential2} as in Ref.~\cite{Akrami:2017cir}, so that the equation of state of the scalar field becomes $-1$ today. Note however that the construction presented here does not provide an explanation of the magnitude of the offset, i.e. does not address the cosmological constant problem.

\begin{figure}[h!]
	\centering
	\includegraphics[width=.85\columnwidth]{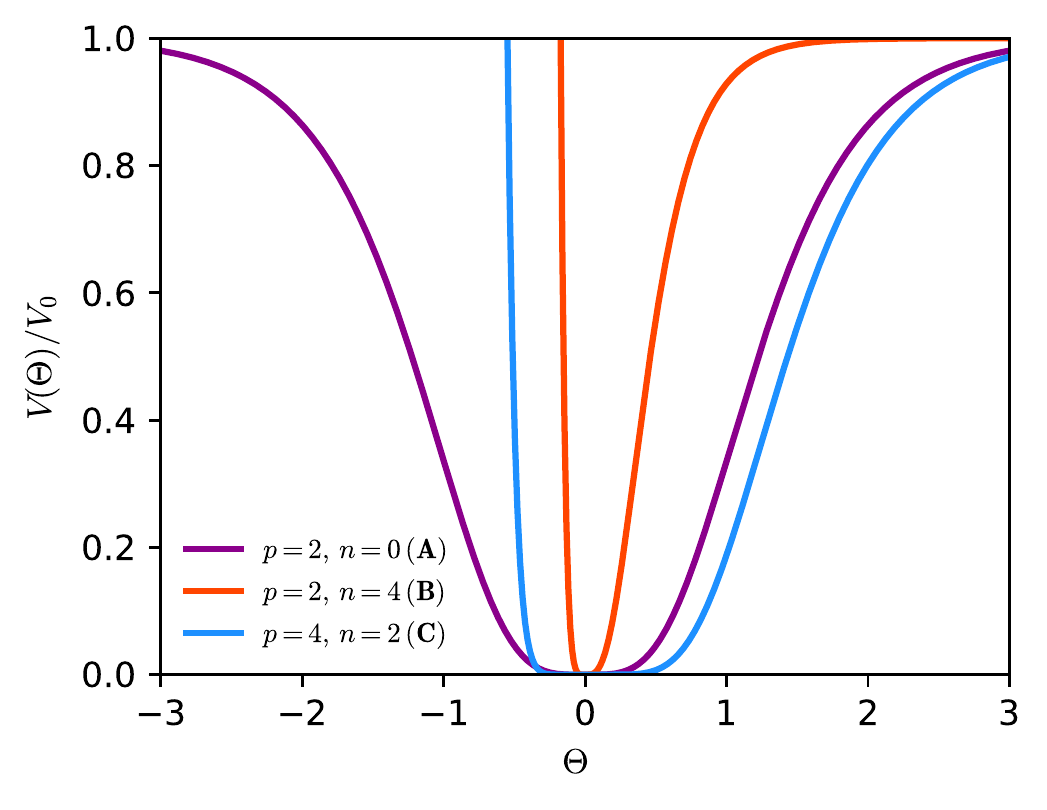}
	\caption{We plot the potential in Eq.~\eqref{eq:potential2} for $(p,\,n)=\{(2,\,0),\,(2,\,4),\,(4,\,2)\}$.    }
	\label{fig:Potential}
\end{figure}

We show the potential for three particular choices of $(p,\,n)=\{(2,\,0),\,(2,\,4),\,(4,\,2)\}$, that we label $\{${\bf A}, {\bf B}, {\bf C}$\}$ respectively, in Fig~\ref{fig:Potential}.\footnote{Since the purpose of this paper is to show that Early Dark Energy models can be incorporated in the $\alpha$-attractor framework, we have only restricted ourselves to three representative models $\{${\bf A}, {\bf B}, {\bf C}$\}$ which constitute a small sample of the full parameter space described by the potential in Eq.~\eqref{eq:potential2}. We stress, however, that our choice is not the only possible one and other choices can lead to results similar to the ones shown in this paper.} The reason for this  choice will become clear in the following. Note that the potential is asymmetric around the origin in the last two cases for which $n\neq0$. As we will see,  the potential in Eq.~\eqref{eq:potential2} captures all the interesting phenomenological EDE models, but other functional forms can in principle be chosen according to Eq.~\eqref{eq:potential1}.
    
We now discuss the cosmological evolution of $\alpha$-attractor EDE. 
The dynamics of the scalar field is similar to other models of EDE studied in the literature and is essentially that of an ultralight axion field \cite{Marsh:2015xka}. The scalar field starts from its initial value $\Theta_i$ deep in the radiation era and remains frozen because of the Hubble friction. The energy density of the scalar field is subdominant in this regime and its equation of state $w_{\rm EDE}\equiv P_{\rm EDE}/\rho_{\rm EDE}$ is equal to $-1$, hence the name ``Early Dark Energy".  Eventually, the effective mass of the scalar field becomes comparable to the Hubble rate $H$ and $\phi$ starts to thaw. The redshift $z_c$ at which this occurs can be implicitly defined from the relation $\frac{\partial^2 V(\phi_i)}{\partial\phi^2}\simeq 9 H^2(z_c) $ \cite{Marsh:2015xka}. After $z_c$, the Hubble friction is too weak to keep the scalar field up its potential and it rolls down in a very short time. When this happens, the potential energy of the scalar field is converted into a kinetic one and a certain amount of energy, parameterized by $f_{\rm EDE}\equiv\rho_{\rm EDE}(z_c)/3 M_{\rm pl}^2 H^2(z_c)$ is injected into the cosmic fluid. Depending on the slope of the potential and its structure around the minimum, the scalar field then starts to oscillate or simply freezes again once it has exhausted its inertia. The critical redshift $z_c$ and the value of the energy injection $f_{\rm EDE}$ are the key parameters describing all EDE models \cite{Poulin:2018dzj}. 
As we are going to discuss, the  shape of the energy injection and $w_{\rm EDE}$ crucially depend on the different possible dynamics of the scalar field after $z_c$. The scalar field energy density quickly redshifts away after $z_c$ and its contribution becomes subdominant with respect to the other components of the Universe.
 
\begin{figure}[h!]
	\centering
	\includegraphics[width=.32\columnwidth]{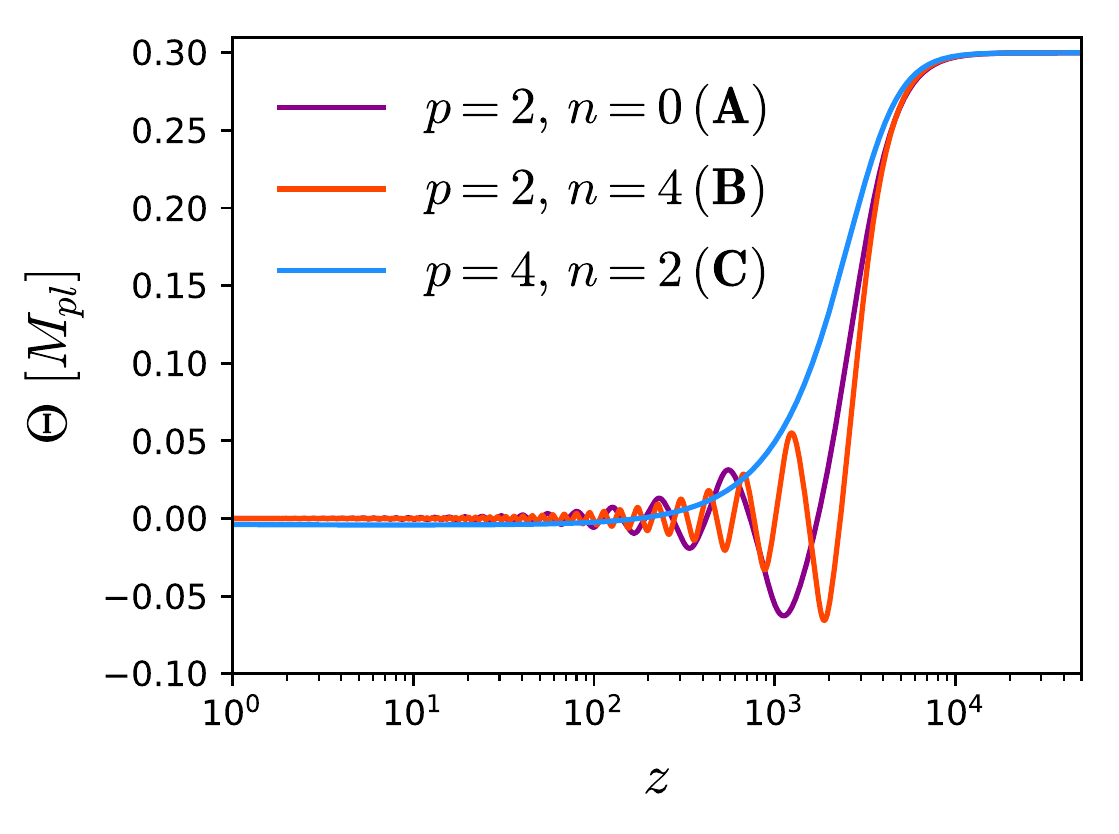}
	\includegraphics[width=.32\columnwidth]{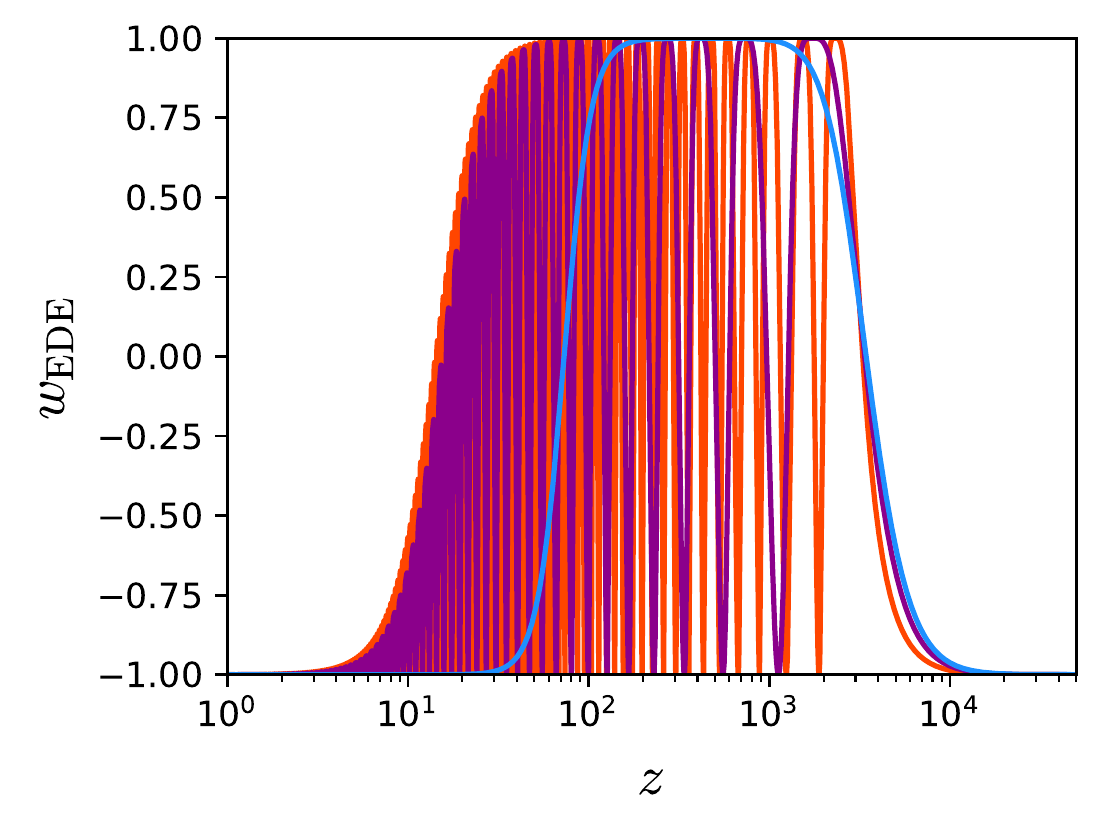}
	\includegraphics[width=.32\columnwidth]{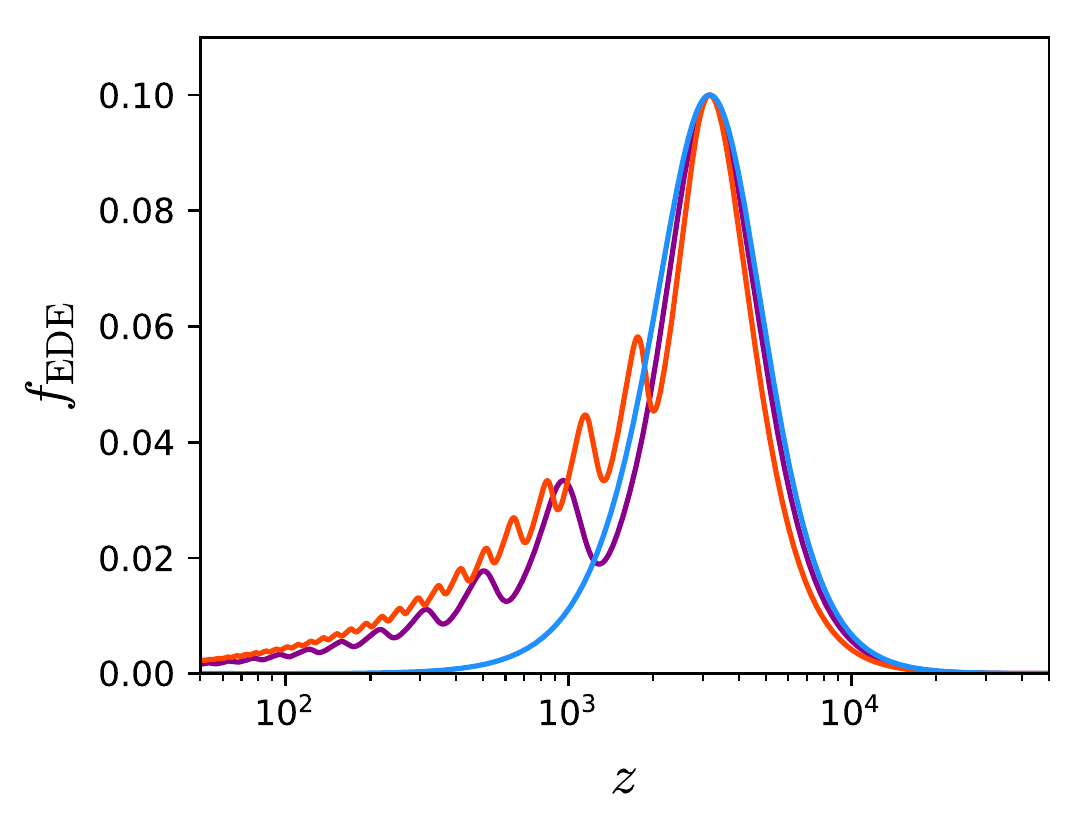}
	\caption{We plot the evolution of the normalized scalar field $\Theta$ [Left], equation of state parameter $w_{\rm EDE}$ [Center] and the energy injection $f_{\rm EDE}$ [Right] for the three models  with $(p,\,n)=\{(2,\,0),\,(2,\,4),\,(4,\,2)\}$. For definiteness, we have chosen $f_{\rm EDE}=0.1$, $\log_{10} z_c=3.5$ and $\Theta_i=0.4$.}
	\label{fig:Background}
\end{figure}

We show in Fig.~\ref{fig:Background} the EDE dynamics for the three ({\bf A}, {\bf B} and {\bf C}) cases mentioned above. In particular, we plot the scalar field evolution, its equation of state and the energy injection in the left, central and right panels respectively (see the caption for the parameters used).

\onecolumngrid
    
\twocolumngrid
    
In the cases {\bf A} and {\bf B }, the scalar field oscillates at the bottom of its potential leading to a highly oscillatory equation of state. In the {\bf A} case,  the potential is $\tanh^4 \Theta\sim \Theta^4$ around $\Theta\simeq0$ and therefore the shape for the energy injection closely resembles the one obtained  in the so-called rock'n'roll model of Ref.~\cite{Agrawal:2019lmo} where $V(\phi)\propto\phi^4$. On the other hand, the {\bf B} case looks more similar to the original EDE proposal of Ref.~\cite{Poulin:2018cxd} (see e.g. Fig.~2 of Ref.~\cite{Smith:2019ihp}). However, given the asymmetry of our potential for the {\bf B} case, the oscillatory pattern in the energy injection shows an asymmetric amplitude of odd and even peaks in the oscillations. Although this is barely visible in Fig.~\ref{fig:Background}, this effect is more pronounced for larger $\Theta_i$ and  might in principle lead to distinct results, as the CMB power spectrum is very  sensitive to the shape of $f_{\rm EDE}(z)$ \cite{Knox:2019rjx}. Indeed, because of such a sensitivity, the oscillatory patterns of the scalar field in models {\bf A} and {\bf B } leave different imprints on the CMB angular spectra as shown in Refs.~\cite{Agrawal:2019lmo} and \cite{Smith:2019ihp}. Therefore, although at a first glance their background evolution might look similar, it is important to explore the phenomenology of both of them separately.

The case {\bf C} is instead different. Unlike the first two oscillatory models, for this choice of $p$ and $n$, the bottom of the potential is very close to flat and the scalar field shows no oscillations. As anticipated, this model looks indeed similar to the canonical Acoustic Dark Energy (cADE) model proposed in Ref.~\cite{Lin:2019qug}. As in cADE (see also Ref.~\cite{Alexander:2019rsc}), the potential energy is suddenly converted to a kinetic one and the scalar field remains in a \emph{kination} regime in which $w_{\rm EDE}=1$, and its energy is kinetically  dominant until it redshifts away. However, differently from cADE, where the potential was introduced by patching a quartic potential for positive values of $\phi$ to $V(\phi)=0$ for negative ones, our potential {\bf C}  is consistently embedded in the $\alpha$-attractor's construction.

Although we have focused on these three specific cases that well reproduce some cases in the literature, we stress that other possibilities can be obtained for other combinations of the potential parameters $(p,\,n)$. 

\section{Cosmological constraints and implications for the $H_0$ tension} \label{sec:MCMC}

In this Section, we perform a Markov-Chain Monte Carlo (MCMC) analysis with cosmological data and investigate the capability of $\alpha$-attractor EDE models to ease the $H_0$ tension. We use the publicly available code {\tt MontePython-v3}\footnote{\href{https://github.com/brinckmann/montepython\_public}{https://github.com/brinckmann/montepython\_public}} 
\cite{Audren:2012wb,Brinckmann:2018cvx} interfaced  with our modified version of {\tt CLASS}\footnote{\href{https://github.com/lesgourg/class\_public}{https://github.com/lesgourg/class\_public}} 
\cite{Lesgourgues:2011re,Blas:2011rf}

We include several datasets in our analysis. 
We consider CMB measurements  from the {\em Planck} 2018 legacy release (P18) on temperature,  
polarization, and weak lensing CMB angular power spectra \cite{Aghanim:2019ame,Akrami:2018vks}. 
The high-multipole likelihood $\ell \geq 30$ is based on the {\tt Plik} likelihood.
We use the low-$\ell$ likelihood combination at $2 \leq \ell < 30$:
temperature-only {\tt Commander} likelihood plus the {\tt SimAll} EE-only likelihood.
For the {\em Planck} CMB lensing likelihood, we consider the conservative multipole range, i.e. $8 \le \ell \le 400$.
To provide late-time information, complementary to the CMB anisotropies, we use  the Baryon Spectroscopic Survey (BOSS) DR12 \cite{Alam:2016hwk}
``consensus" results on baryon acoustic oscillations (BAO) in three redshift slices with effective redshifts $z_{\rm eff} = 0.38,\,0.51,\,0.61$ 
\cite{Ross:2016gvb,Vargas-Magana:2016imr,Beutler:2016ixs}.   Additionally, we use the Pantheon supernovae dataset \cite{Scolnic:2017caz}, which includes measurements of the luminosity distances of 1048 SNe Ia in the redshift range $0.01< z <2.3$.
Finally, we make use of a Gaussian prior based on the determination of the Hubble constant from from Hubble Space Telescope (HST) observations, i.e. $H_0 = 74.03 \pm 1.42$ \cite{Riess:2019cxk}.

We study the cosmological models denoted by {\bf A}, {\bf B} and {\bf C} introduced in the previous section. We sample the cosmological parameters $\{\omega_b,\,\omega_{cdm}, \,\theta_s,\,\ln 10^{10}A_s,\,n_s,\,\tau_\textup{reio},\,f_{\rm EDE},\,\log_{10}\,z_c,\,\Theta_i\}$ using a Metropolis-Hastings algorithm. We consider the following flat priors  on the EDE parameters:  $f_{\rm EDE}\in[10^{-4},\,0.4]$, $\log_{10}\,z_c\in[2.9,\,4.2]$ as in other EDE studies, see e.g. Ref.~\cite{Smith:2019ihp}, and $\Theta_i\in[0.05,\,1.4]$. We have tested that larger priors on $\Theta_i$ give the same results. We consider the chains to be converged using the Gelman-Rubin criterion $R-1<0.01$ \cite{Gelman:1992zz} and adopt the Planck convention for modeling free-streeming neutrinos as two massless species and one massive one with $M_\nu=0.06$ eV. 

Concerning the linearized perturbations, we impose adiabatic initial conditions and solve the exact evolution of the scalar field perturbations $\delta\phi(k,\,\tau)$ in the synchronous gauge \cite{Ma:1995ey}. 

\begin{table*}[h!]
	{\small
		\centering
		\begin{tabular}{|l||c|c|c|c|}
			\hline
			\hline &$\Lambda$CDM
			& $p=2,\,n=0\,({\bf A})$ & $p=2,\,n=4\,({\bf B})$ & $p=2,\,n=4\,({\bf C})$ \\
			\hline
			$10^{2}\omega_{\rm b}$                        & $2.255\pm 0.014$        &  $2.274\pm 0.018$  &  $2.266\pm 0.018$&$2.283\pm 0.024$ \\
			$\omega_{\rm c}$                        &  $0.11854\pm 0.00093 $                   &$0.1265\pm 0.0036$    &$0.1286\pm 0.0041$&$0.1250\pm 0.0038$ \\
			$100*\theta_{s }$             & $1.04205\pm 0.00028$   & $1.04154\pm 0.00037 $   & $1.04134\pm 0.00039$& $1.04161\pm 0.00040$\\
			$\tau_\textup{reio }$                               &$0.0603\pm 0.0076$   &  $0.0602^{+0.0070}_{-0.0081}$  &  $0.0604^{+0.0068}_{-0.0082}$&$0.0583^{+0.0070}_{-0.0079}$\\
			$\ln \left(  10^{10} A_{\rm s} \right)$ &$3.055\pm 0.015$   & $3.067\pm 0.016$   & $3.071\pm 0.016$&$ 3.064\pm 0.015$\\
			$n_{\rm s}$                             &  $0.9701\pm 0.0037$ &  $0.9803\pm 0.0057 $  &$0.9795\pm 0.0054$& $ 0.9797\pm 0.0063$ \\
			\hline
			$\log_{10}\,z_c$                        &  $-$  &  $3.550^{+0.074}_{-0.061}$  &  $3.510^{+0.044}_{-0.053}$ &$3.528^{+0.058}_{-0.10}$\\
			$f_{\rm EDE}$                        &  $-$  &  $0.065\pm 0.026$  &  $0.082\pm0.029$&$0.048^{+0.029}_{-0.024}$ \\
			$\Theta_i$                      &$-$  &  $< 0.554$  &  $< 0.184$ &$< 0.322$\\
			$\log_{10} \alpha$                        &  $-$  &  $ -0.3^{+1.4}_{-1.1}$  &  $0.65^{+1.5}_{-0.99}$ &$0.0^{+1.6}_{-2.6}$\\
			$\log_{10}\,V_0/{\rm eV}^4$                       &  $-$ &  $1.7^{+2.5}_{-1.7}$&$1.4^{+2.4}_{-1.7}$     &$3.4^{+4.5}_{-4.0}$\\
															\hline
			$H_0$ [km s$^{-1}$Mpc$^{-1}$]             &  $68.29\pm 0.42$     &$70.28\pm 0.94$    & $70.9\pm 1.1 $  &$69.88\pm 0.99$\\
			$\sigma_8$                              & $0.8105\pm 0.0064 $ &  $0.8255\pm 0.0090$  &$0.8271\pm 0.0091$ &$0.829\pm 0.012$\\
			$r_s$ [Mpc]                             &  $147.21\pm 0.23$ & $143.0\pm 1.8$  &     $142.0\pm 2.0$&$143.7^{+2.5}_{-2.2}$\\
			\hline
			$\Delta \chi^2_\textup{min}$                         &  & $-9.68$ & $-10.36$ &$-8.66$\\
			\hline
			\hline
	\end{tabular}}
	\caption{\label{tab:MCMC} 
		Constraints on main and derived parameters of the three examples in the main text considering 
		Planck 2018 data (P18), BAO, Pantheon and SH0ES data.  We report mean values and the 68\% CL, except for the subset of EDE parameters $\{\,\Theta_i,\,\log_{10}\, V_0/{\rm eV}^4,\,\log_{10}\, \alpha \}$ for which we report the 95\% CL.}
\end{table*}

\onecolumngrid

\begin{figure}[h!]
	\centering
	\includegraphics[width=.85\columnwidth]{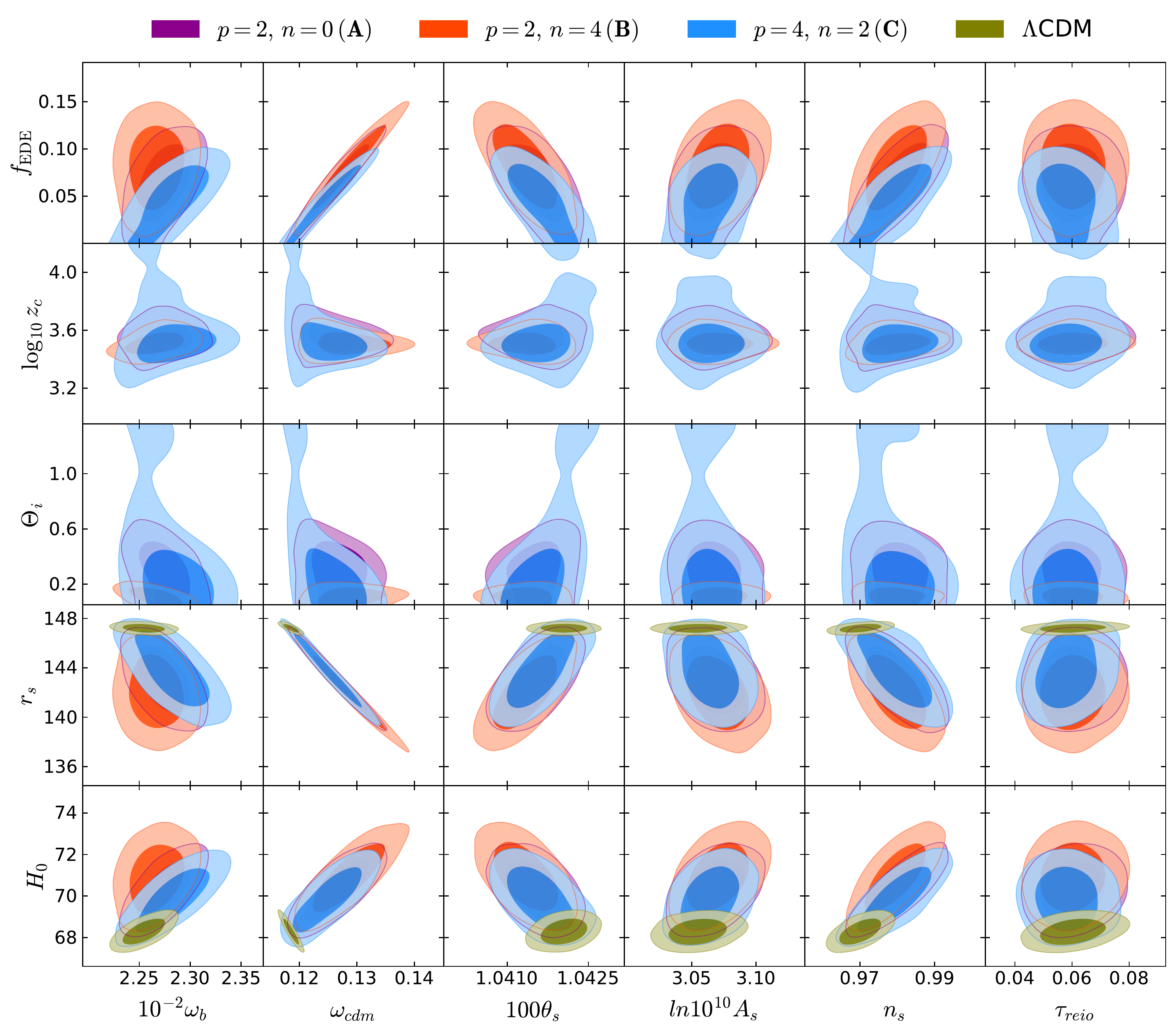}
	\caption{Constraints on main parameters and $H_0$ of the $\alpha$-attractor models {\bf A}, {\bf B} and {\bf C} from Planck 2018 data (P18), BAO, Pantheon and SH0ES data. Parameters on the bottom axis are the standard cosmological parameters, and parameters on the left axis are the EDE parameters that we sample with flat priors, $r_s$ in [Mpc] and  $H_0$ in  [km s$^{-1}$Mpc$^{-1}$]. Constraints for the $\Lambda$CDM model obtained with the same dataset are also shown. Contours contain 68\% and 95\% of the probability.}
	\label{fig:MCMC}
\end{figure}
\twocolumngrid

\begin{figure}[h!]
	\includegraphics[width=.85\columnwidth]{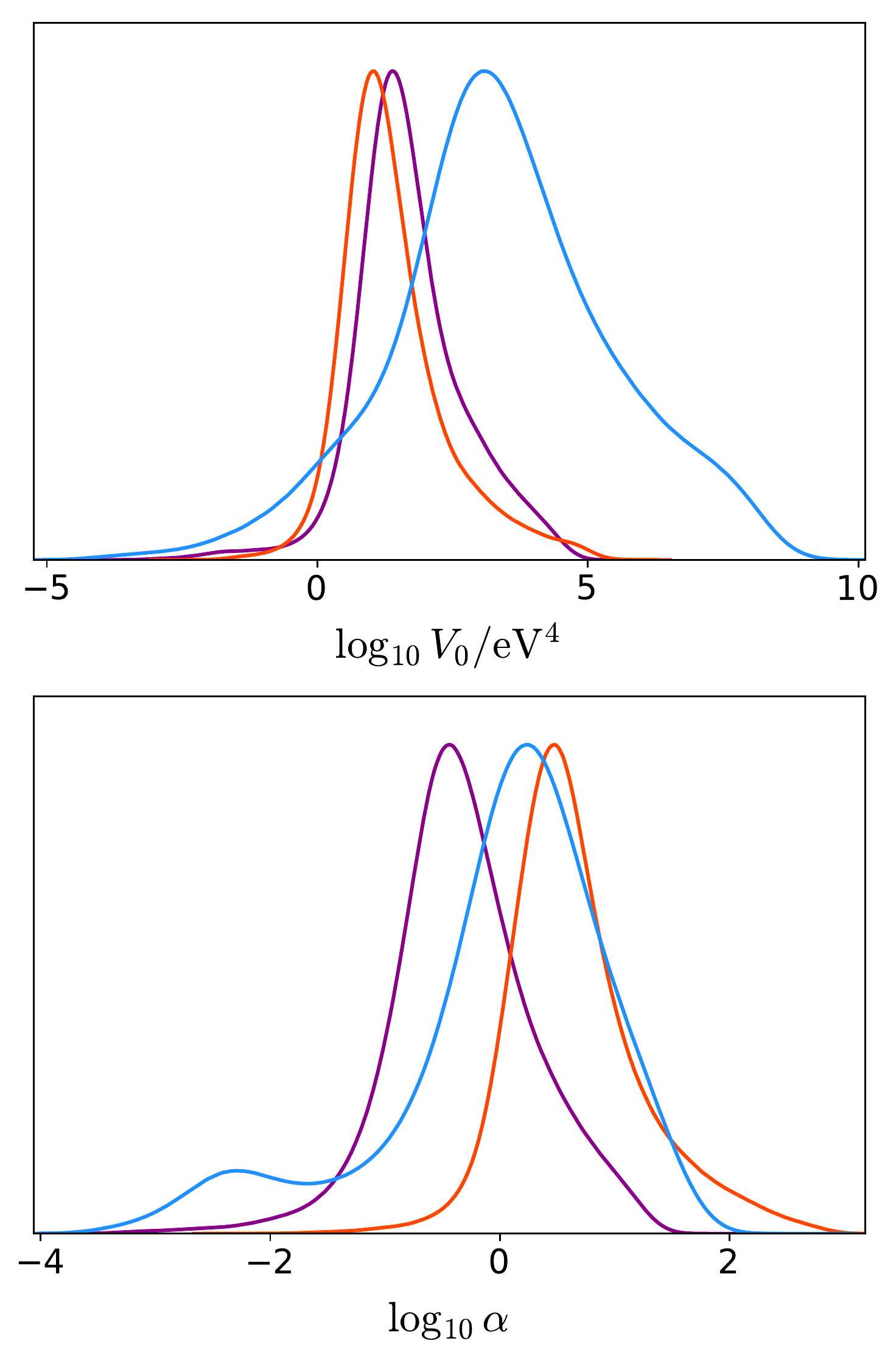}
	\caption{One dimensional derived posterior distribution of the potential parameters $\log_{10}\, V_0/{\rm eV}^4$ and $\log_{10}\, \alpha$. The convention for the colors used is the same of Fig.~\ref{fig:MCMC}.}
	\label{fig:MCMC1d}
\end{figure}

Our results are summarized in Table~\ref{tab:MCMC}, where we report   the reconstructed mean values and the 68\% and 95\% CL, and Fig.~\ref{fig:MCMC}, which has been obtained using {\tt GetDist}\footnote{\href{https://getdist.readthedocs.io/en/latest}{https://getdist.readthedocs.io/en/latest}} \cite{Lewis:2019xzd}, where we plot the reconstructed two-dimensional posterior distributions of the main and derived parameters.
We also  plot in Fig.~\ref{fig:MCMC1d} the one-dimensional posterior distributions on the parameters of the potential $V_0$ and $\alpha$.

\section{Results}
\label{sec:results}
We now comment the results of the previous Section. As expected, all  three cases lead to larger values for the Hubble parameter $H_0$, as can be seen from Tab.~\ref{tab:MCMC}. We find that the larger energy injection is allowed in the model {\bf B} for which $f_{\rm EDE}=0.082\pm0.029$ results in $H_0= (70.9\pm1.1)$ km s$^{-1}$Mpc$^{-1}$ at 68\% CL. This is followed by model ${\bf A}$ for which $f_{\rm EDE}=0.065\pm 0.026$ results in $H_0= (70.28\pm0.94)$ km s$^{-1}$Mpc$^{-1}$ and model ${\bf C}$ for which $f_{\rm EDE}=0.048^{+0.029}_{-0.024}$ results in $H_0= (69.88\pm 0.99)$ km s$^{-1}$Mpc$^{-1}$. As in other EDE models, there is a clear degeneracy between the comoving sound horizon $r_s$ and $f_{\rm EDE}$ which is responsible for the enhanced expansion rate around the epoch of matter-radiation equality and therefore the lower $r_s$. On the other hand, to successfully preserve the fit to the CMB data a shift to a  larger CDM density $\omega_c$  is needed, leading to the degeneracy between $r_s$ and $\omega_c$ seen in Fig.~\ref{fig:MCMC} and to a worsening of the $\sigma_8$ tension \cite{Hill:2020osr}. 

In all the three models considered, cosmological data require that the initial value for $\Theta_i \sim {\cal O} (1)$ after nucleosynthesis, once the slow-roll regime is considered. These values are typical also for other EDE models. For these values, the scalar field is hung up in the descending slope of the potential shown in Fig.~\ref{fig:Potential} after nucleosynthesis: this range of values does not exclude that the scalar field could have been in the plateau outside the slow-roll regime at the beginning of the relativistic era. 

Our results are in agreement with the comparison between {\bf A}, {\bf B} and {\bf C} and models in the literature made in the previous Section. Indeed, the higher value of $H_0$ in the literature of EDE models can be found in the original EDE proposal of Ref.~\cite{Poulin:2018cxd}. However, due to our use of more recent CMB data and perhaps the slightly asymmetric oscillations in the energy injection, our inferred value for $H_0$ is somewhat lower. In fact Ref.~\cite{Hill:2020osr} also found an $H_0$ similar to ours when analyzing the model of Ref.~\cite{Poulin:2018cxd}, adopting the same dataset used here.

Furthermore, contrary to Refs.~\cite{Agrawal:2019lmo,Lin:2019qug}, the potential in Eq.~\eqref{eq:potential2} contains two free parameters, and despite the similar shape of the energy injection in the models {\bf A} and {\bf C} respectively, the enhanced number of degeneracies between parameters leads to a slightly different inferred $H_0$ also in this case.

We note an interesting difference between these EDE models based on $\alpha$-attractor-like potentials and those  inspired by ultralight axionlike fields as Refs.~\cite{Poulin:2018dzj,Poulin:2018cxd,Smith:2019ihp}. In models involving cosine potentials, the axion decay constant $f$ has to take values of order $\mathcal{O}(M_\textup{pl})$ to solve the $H_0$ tension \cite{Hill:2020osr}, in contrast with the weak gravity conjecture \cite{ArkaniHamed:2006dz}.  
Our results show instead that the allowed range for $\alpha$ is natural in terms of model building and also includes the discrete values for $\alpha$ motivated by maximal supersymmetry \cite{Ferrara:2016fwe,Kallosh:2017ced}. It is also interesting to note that we have here an apparent non-zero detection of $\alpha$ for EDE, whereas at present we have only upper bound on $\alpha$ for inflationary models \cite{Akrami:2018odb}. Note, however, that in order to correctly claim a non-zero detection of  $\alpha$, we should also vary the potential parameters $n$ and $p$ in our MCMC and marginalize over them.

\section{Conclusions}
\label{sec:conclusions}
	
In this paper we have studied a new model of Early Dark Energy (EDE) consisting of a minimally coupled scalar field in the framework of $\alpha$-attractors. As is typical in EDE models, the scalar field remains frozen during the radiation era, until it becomes massive and quickly rolls down the potential, injecting energy into the cosmic fluid and temporarily enhancing the expansion rate of the Universe. The shape of the energy injection in redshift is crucial to solve the $H_0$ tension and depends on the structure of the potential around its minimum. The only constraint in $\alpha$-attractors is that the potential has to be of the form $V(\phi)= f^2\left[\tanh(\phi/\sqrt{\alpha})\right]$, giving in principle a large freedom to the model building.  

Adopting the simple potential in Eq.~\eqref{eq:potential2} as a working example, we have shown that it is indeed possible for the energy injection to take several different shapes in the single unified framework of $\alpha$-attractors and reproduce results from different studies in the literature. To illustrate this, we have analyzed three example models. In the first two ({\bf A} and {\bf B}) the scalar field oscillates at the bottom of the potential in a way that resembles the works \cite{Agrawal:2019lmo} and \cite{Poulin:2018cxd} respectively. Note however, that our second example slightly differs from~\cite{Poulin:2018cxd} since the asymmetry of the potential around $\phi=0$ leads to an asymmetric pattern of oscillations in the energy injection. In our third model ({\bf C}), instead, the scalar field never oscillates and quickly transfers its potential energy to kinetic one, undergoing a temporary phase of kination, as in Ref.~\cite{Lin:2019qug}.

We have used the latest \emph{Planck} 2018 CMB temperature, lensing and polarization data together with a variety of high and low $z$ BAO measurements, SNe Ia data from Pantheon and the SH0ES estimate of the Hubble constant, and run an MCMC simulation to constrain the model parameters. We have found that all the models can significantly alleviate the $H_0$ tension, the best being the model {\bf B}, for which an energy injection of $f_{\rm EDE}=0.082\pm0.029$ at the redshift $\log_{10}\,z_c=3.510^{+0.044}_{-0.053}$ leads to an inferred value of the Hubble rate today of $H_0= (70.9\pm1.1)$ km s$^{-1}$Mpc$^{-1}$ at 68\% CL.

As noticed in the literature, EDE models change the best-fit cosmological parameters from $\Lambda$CDM such as $n_s$, $A_s$ and $\omega_c$. This could lead to tension with large scale structure observations such as weak gravitational lensing \cite{Hill:2020osr}, although the conclusion depends on the CMB data used in the analysis \cite{Chudaykin:2020acu}. Another interesting consequence is the spectral index $n_s$, which tends to be larger than the one obtained in $\Lambda$CDM, $n_s \sim 0.98$. This has an interesting implication for inflationary models. For example, some inflation models based on $\alpha$-attractors predict a larger $n_s$ if reheating occurs gravitationally \cite{Dimopoulos:2017zvq,Akrami:2017cir}. These $\alpha$-attractor inflation models can be combined with early dark energy models as a two field model or quintessential inflationary models as done in Refs.\cite{Dimopoulos:2017zvq,Akrami:2017cir}. It will be interesting to explore this possibility and revisit inflation models in the light of new constraints on $n_s$ \cite{Martin:2013tda, Vennin:2015vfa,Akrami:2018odb}.

\vspace{0.5cm}
\noindent
\textbf{Note added:}
After this paper was completed, new studies involving LSS data appeared on the arXiv. The results obtained in Ref.~\cite{Hill:2020osr} that first explored to which extent the shift towards larger values of $\omega_c$ in EDE models can be tolerated by weak lensing and redshift-space distortions data, have been confirmed in more recent papers that used the full-shape of the power spectrum \cite{Ivanov:2020ril,DAmico:2020ods,Bull:2020cpe}. The conclusions of these papers, which do not include the SH0ES data in the analysis, is that LSS data break the degeneracies between the EDE and cosmological parameters mentioned in Sec.~\ref{sec:results} and EDE models (which predict in general a higher sigma8) are not able to significantly ease the $H_0$ tension. More optimistic results are found in Ref.~\cite{Klypin:2020tud}.

 
\begin{acknowledgments}
\end{acknowledgments}
MB acknowledges the Marco Polo program of the University of Bologna for supporting a visit to the Institute of Cosmology and Gravitation at the University of Portsmouth, where this work started. 
FF acknowledges contribution from the contract ASI/INAF for the Euclid mission n.2018-23-HH.0 and by ASI Grant 2016-24-H.0.
AEG and KK received funding from the European Research Council under the European Union’s Horizon 2020 research and innovation programme (grant agreement No. 646702 ``CosTesGrav"). KK is also supported by the UK STFC ST/S000550/1. WTE is supported by an STFC consolidated grant, under grant no. ST/P000703/1. Numerical computations for this research were done on the Sciama High Performance Compute cluster, which is supported by the ICG, SEPNet, and the University of Portsmouth.

\end{document}